\documentclass[english,aps,prl,tightenlines,floats,floatfix,nofootinbib,twocolumn]{revtex4}
\usepackage[T1]{fontenc}
\usepackage[latin9]{inputenc}
\usepackage{wrapfig}
\usepackage{amsmath}
\usepackage{graphicx}
\makeatletter
\usepackage{bm}
\usepackage{graphics}
\usepackage{epsf}
\makeatother
\usepackage{color}

\def\beq{\begin{equation}}
\def\eeq{\end{equation}}
\def\bea{\begin{eqnarray}}
\def\eea{\end{eqnarray}}
\def\beqa{\begin{equation}\begin{array}{l}}
\def\eeqa{\end{array}\end{equation}}
\def\eqlab#1{\label{eq:#1}}
\def\figlab#1{\label{fig:#1}}

\def\Eqref#1{Eq.~(\ref{eq:#1})}

\def\fref#1{\ref{fig:#1}}
\def\Figref#1{Fig.~\ref{fig:#1}}

\def\al{\alpha}
\def\be{\beta}
\def\ga{\gamma} 
\def\de{\delta} \def\De{\Delta}
\def\veps{\varepsilon}  

\def\la{\lambda}

\def\th{\theta}  \def\Th{\Theta}
  
\def\vfi{\varphi}

\def\pa{\partial}

\def\pa{\partial}

\def\nn{\nonumber}

\def\lag{{\mathcal L}}

\def\mathscr{\mathcal}

\def\3d{3-D}

\def\ol#1{\overline{#1}}

\def\s#1{\setbox0=\hbox{$#1$}  
   \dimen0=\wd0     
   \setbox1=\hbox{/} \dimen1=\wd1  
   \ifdim\dimen0>\dimen1   
      \rlap{\hbox to \dimen0{\hfil/\hfil}} 
      #1     
   \else     
      \rlap{\hbox to \dimen1{\hfil$#1$\hfil}} 
      /      
   \fi}      %

\makeatother

\usepackage{babel}

\begin{document}

MKPH-T-10-03
\bigskip

\title{Electromagnetic moments of  quasi-stable particle}

\author{Tim Ledwig}

\author{Vladimir Pascalutsa}

\author{Marc Vanderhaeghen}

\affiliation{Institut f\"ur Kernphysik, Johannes Gutenberg Universit\"at Mainz, D-55099 Mainz, Germany}

\date{\today}


\begin{abstract}
We deal with the problem of assigning electromagnetic moments
 to a quasi-stable particle (i.e., a particle with mass located at particle's decay threshold). 
In this case, an application of a small external electromagnetic field changes the energy in a
non-analytic way, which makes it difficult to assign definitive moments.
 On the example of a spin-1/2 field with mass
$M_{*}$ interacting with two fields of masses $M$ and $m$, we show
how a conventionally defined magnetic dipole moment diverges at
$M_{*}=M+m$. We then show that the conventional definition 
makes sense only when the values of the applied magnetic
field $B$ satisfy  $|eB|/2M_{*}\ll|M_{*}-M-m|$.  
We discuss implications of these results to 
existing studies  in electroweak theory, chiral effective-field theory,  and  lattice QCD.
\end{abstract}


\maketitle


Electromagnetic (e.m.) moments of a particle are determined
through observations of the particle's behavior in an applied electromagnetic field. For example,
the magnetic moment is measured by observing the spin precession
in a magnetic field. In doing so, one assumes that the uniform magnetic field $\vec B$
induces a linear response in the energy:
\beq
\eqlab{linear}
\De E= -\vec \mu \cdot \vec{B},
\eeq 
with $\vec \mu$ being the magnetic moment.
This method works perfectly well 
for stable particles (electron, proton), as well as for many unstable particles 
(muon, neutron, etc.),
which live long enough for their spin precession to
be observed.
In this letter we examine the case 
of a ``quasi-stable" particle, i.e., a particle
with mass $M_*$ that could decay into two (for simplicity) particles with masses
$M$ and $m$, such that
\beq
\eqlab{approximate}
M_* = M+m\,.
\eeq
It turns out that applying the magnetic field in this situation does not lead to a
polynomial energy shift but to a response which is
non-analytic in $B$, typically $\De E\sim |\vec{B}|^{1/2}$.
The square-root behavior is characteristic for the particle-production cut. 
In a more general situation, when $M_* \approx M+m$, 
a polynomial expansion in $B$ can be made as long as
\beq
  \eqlab{cond}
|\vec B| \times [\mathrm{magneton}] \ll  |M_*-M-m|,
 \eeq  
 which thus becomes a condition for the magnetic moment to be observable.

We do not yet know of examples in nature where the masses of 
particles would be tuned to such an extent
that the condition \Eqref{cond} would be violated. For example, the neutron mass is less than
 1 MeV above
the threshold ($M_n-M_p-m_e\approx 0.8$ MeV), but this number is huge when compared to any reasonable value of the magnetic field measured in units of nuclear
magneton: $\mu_N \simeq 3 \times 10^{-14}$ MeV/Tesla. Nevertheless, situations where
the condition \Eqref{cond} is violated are sometimes encountered in theoretical studies.
In the studies of the $W$-boson's magnetic and quadrupole moments as a function of
bottom- and top-quark masses $m_b$ and $m_t$, a singularity at $m_b +m_t= M_W$ 
arises from the $b\overline{t}$ (or $t\overline{b}$) loop contributions. This
singularity was reported firtstly in
\cite{CoutureNG:WWphoton,Argyres:WWphoton} at a time when the value
of $m_t$ was not known yet. 
In lattice Quantum Chromodynamics (QCD), 
the e.m.\  moments of hadron are computed for
various values of light quark masses and, as calculations based on
chiral perturbation theory  show,  cups and 
singularities arise too \cite{Pascalutsa:2004je,Jiang:2009jn}.
In this work we find that the singularities arise in the region where 
the electromagnetic moments are ill-defined, because the condition
\Eqref{cond} is not satisfied.

Our findings are best demonstrated on a simple toy model of three fields: a scalar $\vfi$ and two
Dirac spinors $\psi$ and $\it\Psi$, interacting via the Yukawa type of coupling:
\beq
\lag_{\mathrm{int}} = g\, \Big(\, \ol{\it\Psi} \,\psi\,\vfi  + \ol\psi \,{\it \Psi}\, \vfi^\ast \Big) ,
\eeq 
with $g\ll 1$, a small coupling constant. We denote the masses of $\vfi$,
$\psi$ and $\it\Psi$ respectively as: $m$, $M$, and $M_\ast$, and will later on
focus on the region specified
by \Eqref{approximate}. 

Suppose the field $\it\Psi$, as well as one of the
other two fields, has an electric charge $e$, and couples minimally to
electromagnetism.  We look for its anomalous magnetic moment (a.m.m.) $\kappa_\ast$ at
leading order in the coupling $g$. 
\begin{figure}[t]
\begin{center}
\includegraphics[scale=0.35]{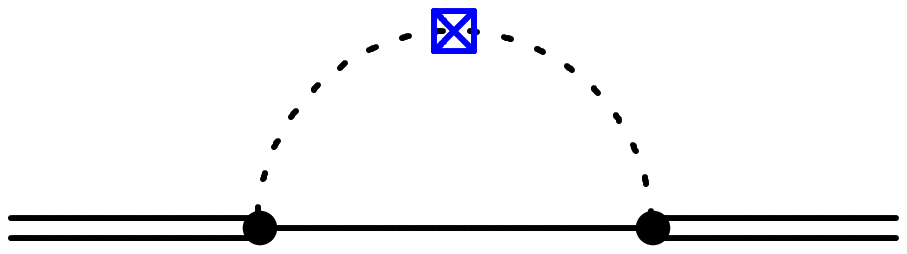}
\hskip8mm
\includegraphics[scale=0.35]{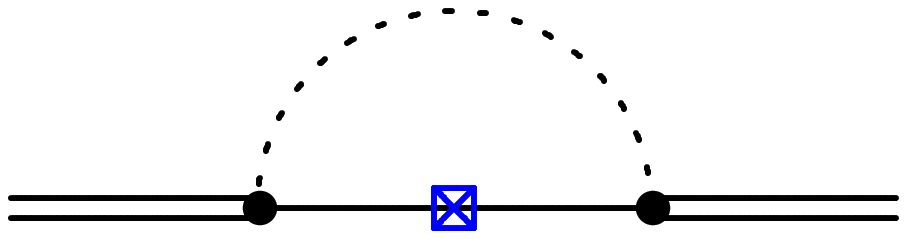}
\end{center}
\caption{One-loop electromagnetic vertex corrections.
Double lines, single and dotted lines denote the propagators of  $\it\Psi$, $\psi$, and $\vfi$, respectively. 
Dots denote the Yukawa coupling and  rectangles the minimal electromagnetic coupling.}
\figlab{diag}
\end{figure}
Depending on whether $\vfi$ or $\psi$ is charged we ought to consider the electromagnetic vertex
corrections shown in \Figref{diag}, and obtain (unprimed: $\vfi$ charged, $\psi$ neutral, 
or primed:  $\psi$ charged, $\vfi$ neutral):
\bea
\eqlab{kappa1}
\kappa_\ast &= & \frac{2 g^2}{(4\pi)^2}\int_0^1 \! dx  \,
\frac{- (r+x)\,x (1-x) }{x \mu^2 - x (1-x) + (1-x) r^2},\\
\kappa_\ast^{\prime} &= & \frac{2 g^2}{(4\pi)^2}\int_0^1 \! dx  \,\frac{(r+ x
  ) (1-x)^2 }{x \mu^2 - x (1-x) + (1-x) r^2},
\eea
where $r=M/M_\ast$, $ \mu=m/M_\ast $.

We have checked that for $M_\ast=-M=M_N$ and $m=m_\pi$
being respectively the mass of the nucleon and the pion, 
these expressions reproduce results of the meson theory
(the same result  also arises in chiral perturbation theory at next-to-leading order
\cite{Pascalutsa:2004ga}). 
The minus sign in front of $M$ appears due
to the pseudo-scalar nature of pion.

At $M_\ast = m+M$ (or, $1=\mu+r$), the denominator in the integrands takes the form
$[ x\mu - (1-x)r]^2$, which leads to an essential singularity in these expressions for
any positive $\mu$ and $r$. This can explicitly be seen, for instance, in \Figref{diverge}
where $\kappa_\ast$ is plotted as a function of $\mu$.
If $\vfi$ is pseudo-scalar, $M$ flips the sign in these expressions, and the singularity
is replaced by a cusp.

\begin{figure}[b]
\includegraphics[scale=0.5]{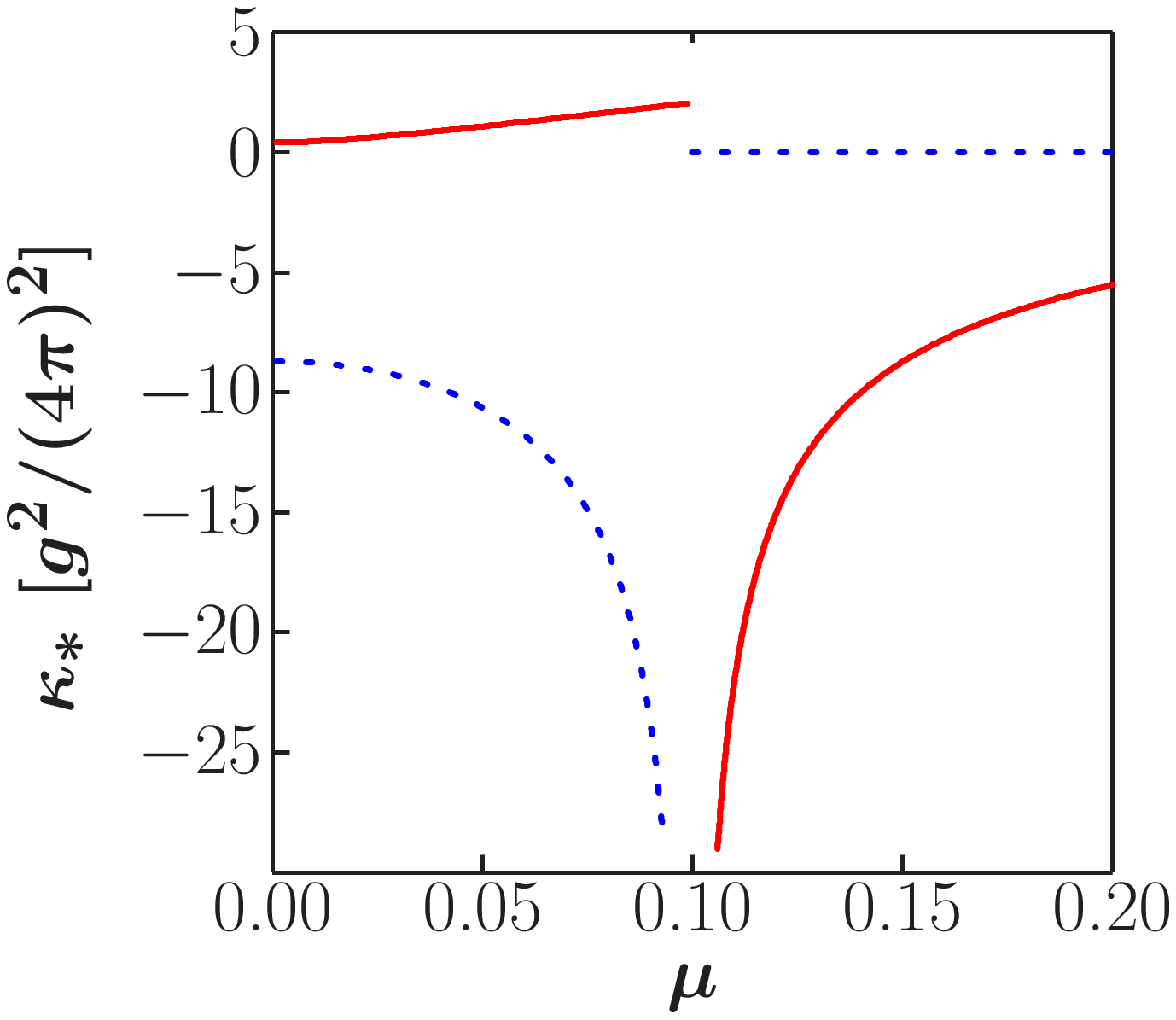}
\caption{The anomalous magnetic moment $\kappa_\ast$ of $\it\Psi$-field as function
of $\vfi$-field mass $\mu$, at fixed value $r=0.9$. 
The red (solid) curve shows the real
part and the blue (dashed) curve the imaginary part of $\kappa$.  (The sign of
the imaginary part is determined by the $i\veps$ prescription.) }
\figlab{diverge}
\end{figure}


The singularity is clearly unphysical, since an infinite value of the magnetic moment
would correspond to a infinite-energy response to an external magnetic field.
To find the correct answer we consider the self-energy of the $\it\Psi$-field in
a constant electromagnetic field, $F_{\mu\nu} \equiv \pa_\mu A_\nu - \pa_\nu A_\mu =\mathrm{const}$. 
Calculations of this sort have been done before, most notably by
Sommerfield and Schwinger~\cite{Sommerfield:electronAMM,Schwinger:book3} 
as a technique to obtain the
correction term of order $\alpha^2_{\textrm{em}}\simeq (1/137)^2$ to the electron's a.m.m..
To cast this technique into a modern field-theoretic language,
we introduce the sources $\it\Th$, $\th$ and $j$ for the fields $\it\Psi$, $\psi$ and $\vfi$,
respectively, and write down the generating functional of the theory,
\bea
Z[ {\it\Th}, \th, j; A] &=& \exp\Big\{-g\! \int d^4 z\,
\Big( \frac{\de^3}{\de j^\ast(z)\, \de\bar\th(z) \, \de{\it\Th}(z) }\nn\\
&+ &  \frac{\de^3}{\de j(z)\,\de\bar{\it\Th}(z)\,\de\th(z) }
\Big)\Big\}\, \exp\Big[ - \!\int d^4 x\, d^4 y\, \nn\\
& \times&   \bar {\it\Th}(x) \,S(x-y;A)\, {\it\Th}(y)  +\ldots \Big],
 \eea
 where 
 \beq
S(x-y;A) = \big[ i \ga^\mu\mbox{$\frac{\pa}{\pa x^\mu}$} - e A_\mu(x)\ga^\mu - M_\ast \big]^{-1} \de^{(4)}(x-y)\,.
\eeq
is the propagator of a charged Dirac particle in the presence of an e.m.\ field.
We then calculate the energy shift induced by the $\it\Psi$-field
self-energy correction in the presence of a constant e.m.\ field. The
dependence on the e.m.\ field comes in the form of the 
$\ga^\mu \ga^\nu F_{\mu\nu}$ structure sandwiched between the free
$\it\Psi$-field states. When the electric contribution is zero, this structure simply yields
the projection of
the magnetic field onto the spin direction. 

The resulting energy-shift, to leading order in $g$, is for the two cases given by:
\bea
\eqlab{shift1}
\De \tilde{E} &=& \frac{ g^2}{(4\pi)^2}\int_0^1 \! dx \,(r+x) \\
&\times& 
\eqlab{shift2}
\ln\Big[1+\frac{ x (1-x)\, \tilde{B}}{x \mu^2 - x (1-x) + (1-x) r^2}\Big],\nn\\
\De \tilde{E}^{\prime} &=& \frac{ g^2}{(4\pi)^2}\int_0^1 \! dx \,(r+x) \\
&\times& 
\ln\Big[1-\frac{ (1-x)^2\, \tilde{B}}{x \mu^2 - x (1-x)  + (1-x) r^2}\Big],\nn
\eea
where the following dimensionless variables are used:
\beq
\tilde{B}=\frac{e B_z}{M_\ast^2}\, ,  \, \quad
\De \tilde{E}=\frac{\De E}{M_\ast} + \frac{1}{2} \tilde B \, ,
\eeq
with $B_z$ the projection of the magnetic field on the spin direction.
The quantity $ \De \tilde{E}$ is the energy shift (in units of $M_\ast$)
due to the a.m.m.\ effect.
In the following we will discuss the unprimed contribution, the primed one can be
obtained analogously.

The e.m.\ field is assumed to be small in comparison with the mass-scale of
particles, and therefore some terms which are higher-order in $\tilde{B}^2$ can 
be neglected. Nevertheless, one can still see that a naive
perturbative expansion in $\tilde{B}$ does not always work. 
In the naive expansion, one finds
\beq
\eqlab{perturb}
\De \tilde{E} = -\frac{\kappa_\ast}{2} \tilde{B}+ \ldots, 
\eeq
with $\kappa_\ast$ given by \Eqref{kappa1}, which recovers the conventional result.
However, around the (in)stability threshold $M_\ast =
M+m$, the naive expansion breaks down, as can be seen from
\Figref{E1combi}  where we plot the energy shift \Eqref{shift1} compared to the
result of the naive perturbative expansion: \Eqref{perturb} with \Eqref{kappa1}. 
It is clear that the two results are very different around the threshold which here is at $\mu=0.1$. 
The size of the region where the two results are different is proportional to the strength of the
magnetic field.

\begin{figure}[tb]
\includegraphics[scale=0.5]{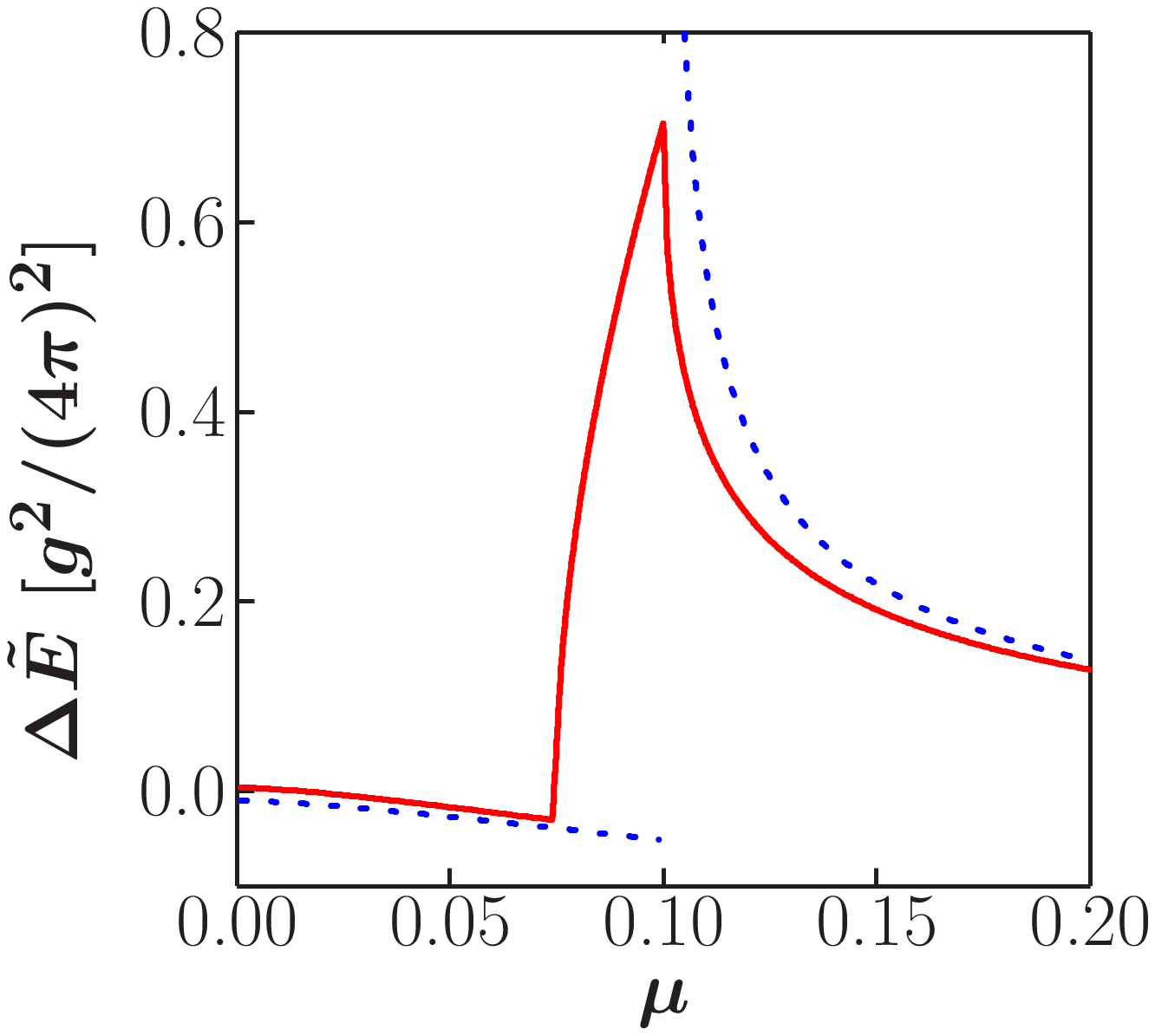}
\caption{The real part of the energy shift $\Delta \tilde{E}$ as
  function of $\mu$ for a fixed magnetic field strength
  $\tilde{B}$. The red (solid) curve is obtained from \Eqref{shift1} while the blue
  (dotted) from \Eqref{perturb} and \Eqref{kappa1}. The parameters
  are chosen $r=0.9$ and $\tilde{B}=0.05$.}
\figlab{E1combi}
\end{figure}

In \Figref{ReE1} we again compare the perturbative and non-perturbative results, 
but now as a function
of the magnetic-field strength. The masses are fixed such that the $\it\Psi$ particle
is stable for solid and long-dashed curves and unstable for medium- and short-dashed curves.
In either situation there is a kink appearing at some value of the magnetic field, which indicates
the crossing over the decay threshold. When $\it\Psi$ is quasi-stable,  $\mu+r=1$, the kink
appears at $B=0$, which makes it impossible to define the moments as derivatives
of the energy response with respect to the e.m.\  field.

\begin{figure}[bt]
\includegraphics[scale=0.5]{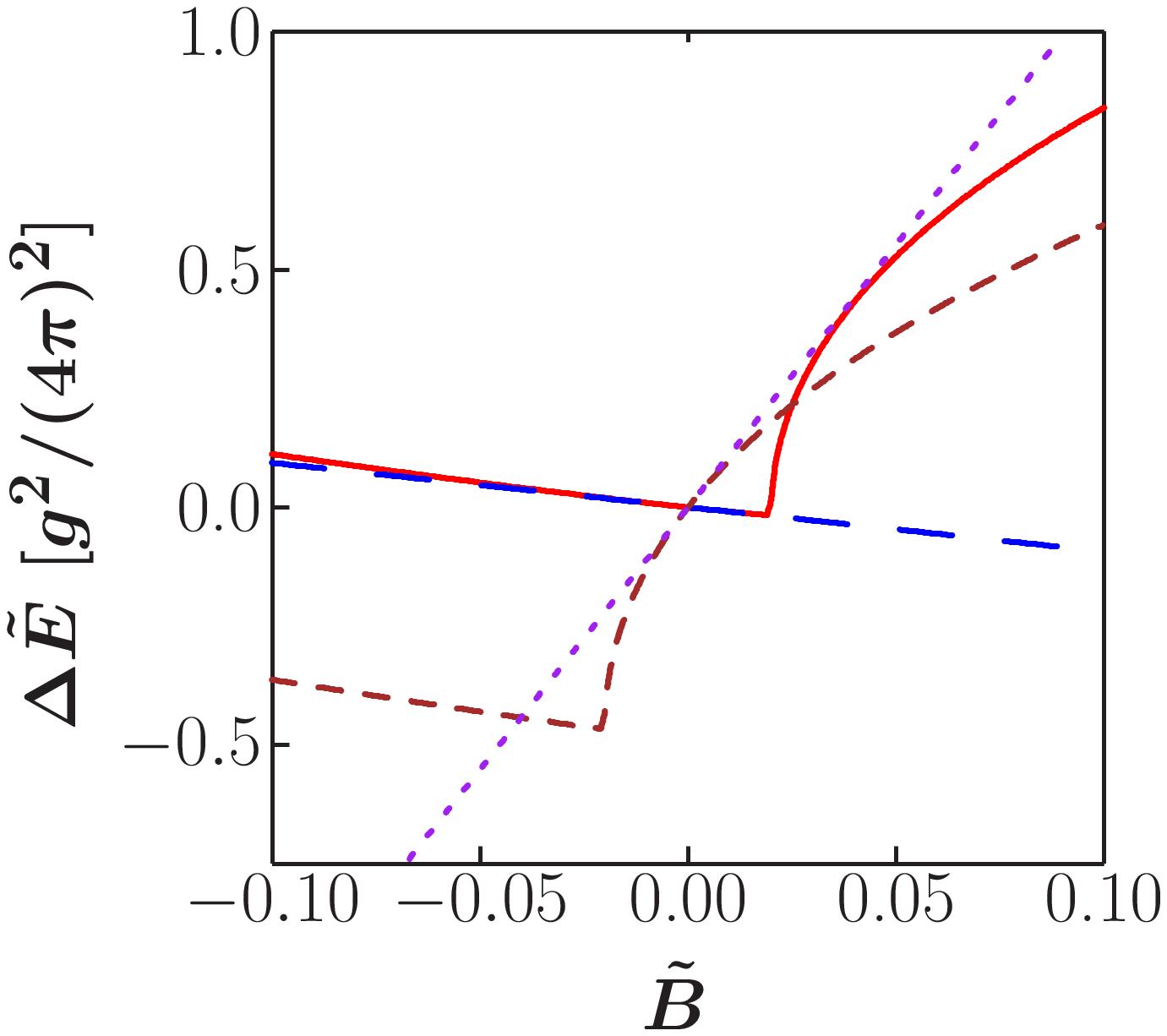}
\caption{The real part of the energy shift $\Delta \tilde{E}$ as function of the magnetic
  field $\tilde{B}$ for a fixed $\mu$. The parameters are chosen as $r=0.9$.
The red (solid) curve is obtained from \Eqref{shift1} and the blue
  (long-dashed) from \Eqref{perturb} with \Eqref{kappa1} for  $\mu=0.09$. The brown
  (medium-dashed) curve is obtained from  \Eqref{shift1} while the purple
(short-dashed) curve from \Eqref{perturb} for  $\mu=0.11$.}
\figlab{ReE1}
\end{figure}

Integration over the Feynman-parameter $x$ in  \Eqref{shift1} 
yields more insight into the non-analytic dependence on the e.m.\ field. The result
can be written as 
\beq
\De \tilde{E} =  \frac{g^2}{(4\pi)^2}\Big\{ (r+\alpha)\, 
({\it \Omega}+\mathcal{A})-[(r+\alpha)\, ({\it \Omega} +\mathcal{A}) ]_{\tilde{B}=0}\Big\}\,\,,
\eeq
 where  $\it\Omega$ is non-analytic in  $\tilde{B}$  :
\bea
{\it \Omega}
&=&\lambda\, \ln\frac{( \alpha +\lambda)(\be+\la) }{(\alpha -\lambda)(\be-\la)},
\eqlab{nonANAE1}
\eea
with 
\begin{eqnarray}
\alpha & = & \frac{1}{2(1-\tilde{B})}\left(1+r^2-\mu^2-\tilde{B}\right),\nn\\
\be & = & \frac{1}{2(1-\tilde{B})}\left(1-r^2+\mu^2-\tilde{B}\right),\\
\lambda & = & \Big[ \alpha^{2}-r^{2}/(1-\tilde{B}) \Big]^{1/2}.\nn
\end{eqnarray}
while the analytic terms are contained in 
\bea
\mathcal{A} &=&-2+\be\ln\mu^2+\alpha\ln r^2
\nn\\
&&
-\,\frac{\mu^2(1-\ln\mu^2)-r^2(1-\ln
    r^2)}{2(\al+r) (1-\tilde{B})}.
\eea
From the expression for $\it \Omega$ we can readily see that a 
Taylor expansion in $B$ only make sense 
when the condition of \Eqref{cond} is satisfied.

The masses of particles are rarely tuned to the
extent that  the condition \Eqref{cond} is in danger. One field of
applications where one does need to pay attention is lattice QCD. 
In modern lattice studies the e.m.\ moments of hadrons can directly be accessed
using the background e.m.\ field method \cite{Bernard:1982yu}.
However, the field strength cannot be arbitrarily small, 
 the periodicity condition poses a lower bound. In the case of magnetic field the bound is:
 $eB\geq 2\pi/(a^2 L) $, or in best case \cite{Damgaard:1988hh}: 
  $eB\geq 2\pi/(a L)^2 $, with length $a$ and integer
 $L$ being respectively
 the lattice spacing and size. For typical modern lattices the lowest possible
 value of the magnetic field can be as large as $10^{14}$ Tesla. Certainly in such strong
 e.m.\ fields the problem raised here becomes relevant and should be studied on a case-by-case basis. 
 
 One typical example would be the case of the $\Delta(1232)$ isobar,
 which magnetic moment has recently been computed using the background
 field method for various pion masses \cite{Lee:2005ds,Aubin:2008hz}. 
 Figure~\fref{condMstar} shows how the condition
 \beq
\left |\frac{eB}{2M_\Delta(M_\Delta-M_N-m_\pi)}\right|\ll1
 \eeq
  can
 be  violated in this type of studies,  but of course for very specific
  values of pion mass and
 the background magnetic field. We  emphasize that  the actual parameters 
 in \cite{Lee:2005ds,Aubin:2008hz},
 do not violate the above
 condition, mainly thanks to the large values of pion mass used in these works. 
 However, current lattice
 calculations begin to approach the pion-mass range where
 this condition would be violated. It would be interesting to see
 how  the non-analytic $B^{1/2}$ behavior emerges in these
 calculations. Of course one can expect this behavior to be 
 shielded by the finite volume effects, the question is to which extent.
 
 \begin{figure}[tb]
\includegraphics[scale=0.5]{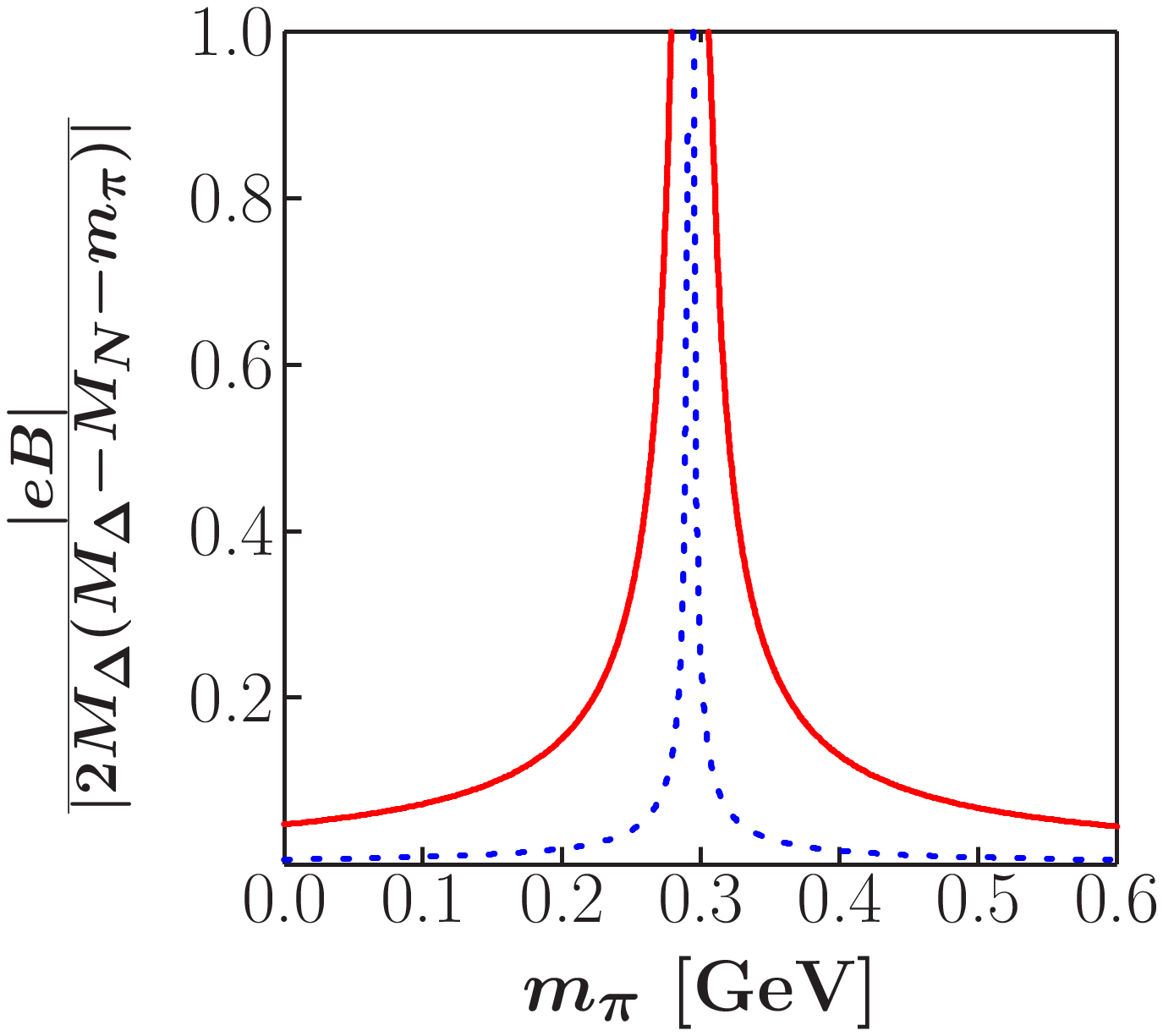}
\caption{The condition \Eqref{cond} for the $\Delta$-nucleon-pion system, $|\frac{eB}{2M_\Delta(M_\Delta-M_N-m_\pi)}|\ll1$, plotted for the range of fields used in~\cite{Lee:2005ds}: $|eB|=0.00108/a^2 \ldots  0.00864/a^2$,
  with $1/a=2\mathrm{\,\,GeV}$ as  function of the pion mass. 
  Red (solid) curve corresponds to the stronger field and the
blue (dashed) to the weaker field. The
  Delta-nucleon mass difference is taken to be constant:  $M_\Delta-M_N=0.293$ GeV.}
\figlab{condMstar}
\end{figure}

 To conclude, the singularities found in calculations of the e.m.\ moments
 of particles, such as W-boson in the Standard Model (prior to the top-quark discovery) or some of the hadrons in chiral effective-field theory, reflect only the limitation of the
calculational technique. When the mass of the particle is near  
 a decay threshold (quasi-stable state), a small external e.m.\ field may induce
 the decay instead of interacting with the particle's e.m.\ moments. 
 We have formulated an exact condition for this effect to occur. In this situation
 an extra care should be taken in defining and determining the moments, as has been described in this work. 
 The present and future lattice QCD calculations of hadron e.m.\ moment
 using the background e.m.\ field technique are a very likely subject to this problem.\\

We acknowledge the support by the Deutsche Forschungsgemeinschaft (DFG).
The work of T.L. was partially supported
by the Research Centre \char`\"{}Elementarkraefte und Mathematische
Grundlagen\char`\"{} at the Johannes Gutenberg University Mainz.

\end{document}